\documentclass[10pt, a4paper]{article}
\usepackage{lrec,multibib,graphicx,tabularx,soul,epstopdf,xstring,pifont}
\usepackage[latin1]{inputenc}

\newcites{languageresource}{Language Resources}

\usepackage{xcolor}

\title{iLCM  - A Virtual Research Infrastructure for Large-Scale Qualitative Data}

\name{Andreas Niekler\textsuperscript{1}, Arnim Bleier\textsuperscript{2}, Christian Kahmann\textsuperscript{1}, Lisa Posch\textsuperscript{2,3},\\ \large\bf Gregor Wiedemann\textsuperscript{1}, Kenan Erdogan\textsuperscript{2}, Gerhard Heyer\textsuperscript{1}, Markus Strohmaier\textsuperscript{4, 2}\\[-5pt],}

\address{\textsuperscript{1}Faculty of Mathematics and Computer Science, University Leipzig, \\
         Augustusplatz 10, 04109 Leipzig, Germany\\
         \{aniekler, kahmann, wiedemann, heyer\}@informatik.uni-leipzig.de\\[4pt]
         \textsuperscript{2}Department of Computational Social Science, GESIS -- Leibniz Institute for the Social Sciences\\
         Unter Sachsenhausen 6-8
50667 Cologne, Germany\\
         \{firstname.lastname\}@gesis.org\\[4pt]
         \textsuperscript{3}Institute of Interactive Systems and Data Science, Graz University of Technology,\\Inffeldgasse 16c, 8010 Graz, Austria\\[4pt]
         \textsuperscript{4}HumTec Institute, RWTH Aachen University\\
         Theaterplatz 14,
52062 Aachen, Germany\\
         markus.strohmaier@humtec.rwth-aachen.de}

\abstract{
The iLCM project pursues the development of an integrated research environment for the analysis of structured and unstructured data in a ``Software as a Service'' architecture (SaaS). The research environment addresses requirements for the quantitative evaluation of large amounts of qualitative data with text mining methods as well as requirements for the reproducibility of data-driven research designs in the social sciences.
For this, the iLCM research environment comprises two central components. First, the Leipzig Corpus Miner (LCM), a decentralized SaaS application for the analysis of large amounts of news texts developed in a previous Digital Humanities project. Second, the text mining tools implemented in the LCM are extended by an ``Open Research Computing'' (ORC) environment for executable script documents, so-called ``notebooks''. This novel integration allows to combine generic, high-performance methods to process large amounts of unstructured text data and with individual program scripts to address specific research requirements in computational social science and digital humanities. \url{ilcm.informatik.uni-leipzig.de} \newline  
}

\begin{document}

\maketitleabstract

\section{Introduction}

Computational social science (CSS) is the interdisciplinary study of socio-cultural phenomena through new kinds of data and technologies. One of its central objectives is the extraction of useful and interpretable knowledge from potentially large behavioral digital datasets. The availability of these datasets has witnessed a fast increase in the past decades through the progressing digitization of social processes. Similar potentials, as well as challenges due to this development, can be observed in fields such as the digital humanities (DH) and communication science (CS).

In CSS, \emph{unstructured} data (usually text) and \emph{structured} data (e.g. metadata, survey data or log/sensor data) are both important sources of information. Through the digitization of social processes, they are available in large quantities. With the iLCM infrastructure, we plan to offer analysis capabilities for reusable and reproducible research with both data types. The manual analysis of qualitative text data is an integral part of the method repertoire of empirical research in various disciplines. The increasing availability of large amounts of digital or retro-digitized text has lead to a growing interest in computer-assisted evaluation methods. Different research methods such as quantitative or qualitative content analysis, discourse analysis or grounded theory methodology can be fruitfully combined with text mining methods \cite{Wiedemann.2016b}. Lexicometric methods such as keyword extraction, frequency- and co-occurrence analysis are already established and widely used in social science text analysis. Machine learning methods such as data-driven clustering of document collections using topic models \cite{Blei2012,Stier.2017a} or the training of automatic classification methods for coding texts \cite{Lemke2015,StierBleier.2018,posch.2015,Wiedemann.2018} begin outreaching into this field. As such applications of text mining help to combine qualitative and quantitative analysis perspectives, they are becoming increasingly relevant as a so-called ``mixed methods'' approach in the social sciences. Related research can be found in political science, sociology, contemporary history research or communication studies. 

Users of computer-assisted text analysis from these disciplines face non-trivial technical and methodological challenges, especially when it comes to the requirement to analyze large amounts of text and reproducibility of research. 
The iLCM infrastructure provides solutions to these challenges. 
\begin{figure}[t]
\begin{center}
\includegraphics[width=\columnwidth]{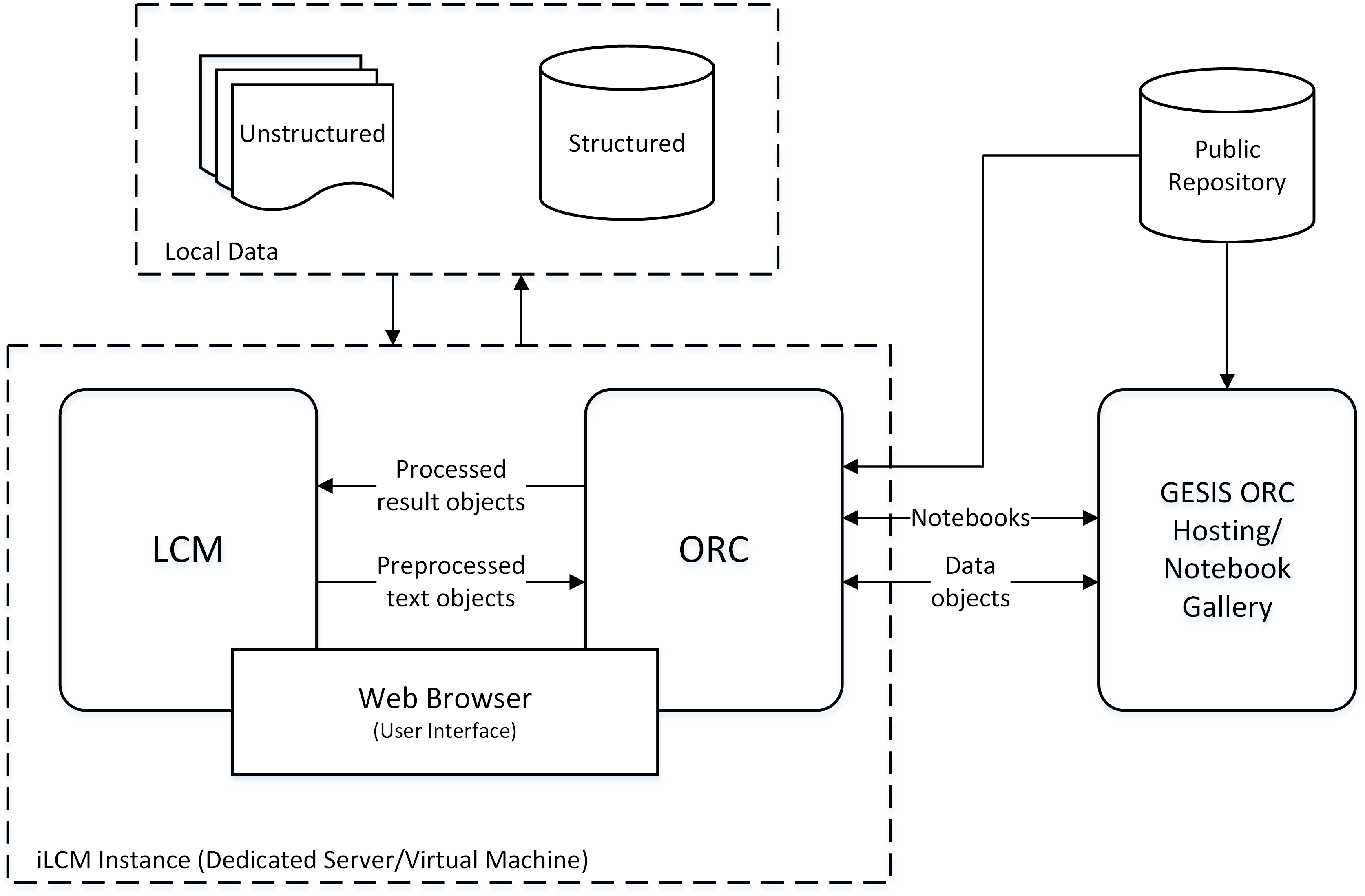} 
\caption{\textbf{iLCM architecture.} The iLCM architecture consists of two major components, the LCM to preprocess and analyze large amounts of unstructured text data, and the ORC component to further process structured data extracted from text or imported from external sources. Analysis scripts on structured data, so-called notebooks, will be published in a central repository.}
\label{fig:architecture}
\end{center}
\end{figure}
As shown in Fig.\ref{fig:architecture}, it consists of two major components. First, the \emph{Leipzig Corpus Miner} (LCM) to \mbox{(pre-)}process and analyze large text collections such as news corpora, social media posts or parliamentary protocols. For this, the LCM provides a browser-based Graphical User Interface (GUI) targeting end-users from applied disciplines who are primarily interested in the application of generic text mining methods on their data. 
Second, the iLCM consists of an \emph{Open Research Computing} (ORC) environment to run scripts in active executable documents, so-called ``notebooks''. 
By this, the ORC component allows analyzing structured data either extracted from texts by the LCM or imported from external datasets. Since the creation of notebooks requires some programming skills, this component is targeted at more experienced end-users who want to realize more complex research designs on top of the generic text mining methods provided by the LCM.
This hybrid architecture enables researchers to process text data in a standardized manner and extend their analyses by highly individualized program scripts that meet the requirements of their specific research tasks. 

In this paper, we present our plans for implementing the iLCM environment and describe its contributions to CSS and DH research. We will cover the capabilities of the Leipzig Corpus Miner (LCM) for text analysis in Section~\ref{sec:lcm} and introduce the idea of Open Research Computing (ORC) for social science applications and the processing of structured data in Section~\ref{sec:orc} In Section~\ref{sec:integration}, we describe how the integration of both components will support the integrated analysis of unstructured and structured data in one environment. In the last sections, we present potential use cases to illustrate the benefits of the iLCM and conclude with a road-map for the ongoing development of the project.

\section{Leipzig Corpus Miner}
\label{sec:lcm}

The LCM component of the environment is based on the previous research project ``ePol'' in which computer scientists together with political scientists created an analysis environment for large diachronic collections of news articles \cite{NWH14,Wiedemann2016}.
This previous project provided a solid basis for an requirements analysis process to design a software providing large scale text mining to social scientists. Relying on the experience gained in several studies conducted by the end-users from the targeted disciplines, we initiated a process of re-implementation to transfer the LCM prototype from the previous project into a universally applicable infrastructure.

The LCM is not a stand-alone program, but rather a server infrastructure comprising a number of components including a document database (MariaDB\footnote{\url{http://mariadb.org}}), an NLP pipeline for preprocessing text data (spaCy\footnote{\url{http://spacy.io}}), a full-text index (Solr\footnote{\url{http://lucene.apache.org/solr}}), a collection of text mining processes (for this, we rely on a selection of mature external packages for the R statistical programming language and additional own implementations), and finally a web application GUI (R Shiny\footnote{\url{http://shiny.rstudio.com}}). 
To make the infrastructure available as a decentralized installation for end-users, it is embedded in a virtual machine ensemble (Docker\footnote{\url{http://www.docker.com}}), which can be easily set up with predefined configuration scripts. 
Docker ensures the executability of the LCM on every system with a running docker environment. All required libraries, software dependencies, and R packages are provided with the docker containers. To set up the LCM container ensemble, we use docker compose\footnote{\url{http://docs.docker.com/compose}} which automates the creation of all involved services in separate containers.
In summary, the LCM integrates components for document management and retrieval, R scripting capabilities for text data, and a GUI for analysis process management and result visualization to enable researchers to conduct text mining on large collections in a systematic manner.

\begin{figure}[!ht]
\begin{center}
\includegraphics[scale=0.125]{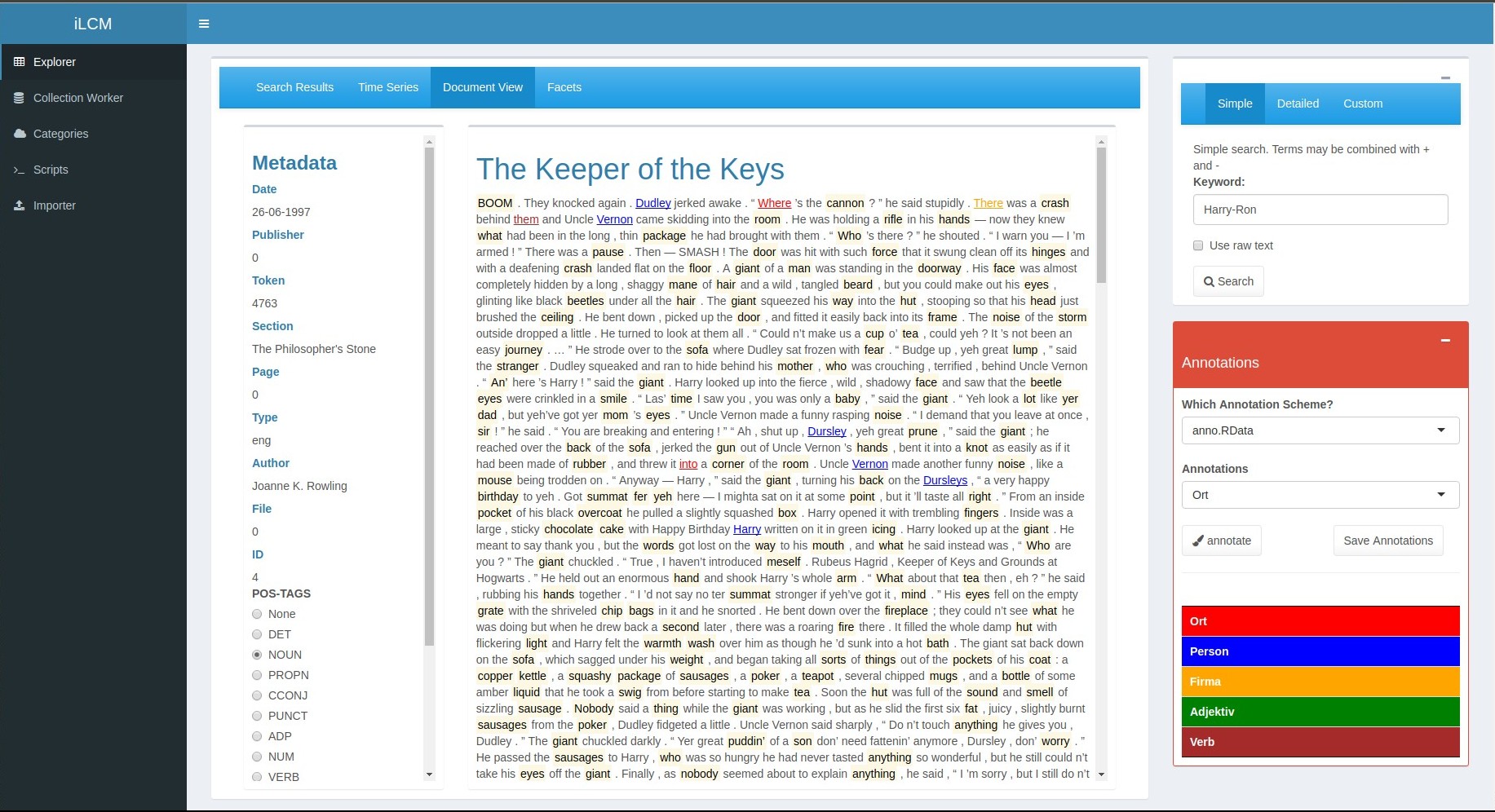} 
\caption{\textbf{LCM Web Application.} Screenshot of user interface (document view).}
\label{iLCM-architecture}
\end{center}
\end{figure}

A screenshot of an early GUI-design is given in Figure~\ref{iLCM-architecture}. It shows the full-text of a retrieved document together with its metadata and some analysis results (keyword highlighting and named entity annotations).  

\subsection{Document Management}

Analyzing large text collections in the context of CSS poses specific requirements to document management. Researchers want to extract information from multiple corpora or compare results of analysis processes with respect to distinct (sub-)collections of documents from one corpus.
Corpora can comprise several million of documents such as series of parliamentary protocols, full volumes of newspapers published over a period of several decades or millions of Twitter tweets regarding a certain hashtag.
To handle such corpora, the LCM provides complex document management and search capabilities.

\paragraph{Corpora:}The LCM allows managing multiple text corpora per user and also allows for sharing of corpora between users. Corpora are not closed at the time of their creation (import) but can be supplemented with additional documents at a later date. This allows the processing of continuous corpora, e.g. collections of newspaper articles, Wikipedia articles or social media data. In order to make the data import as easy as possible for users, import interfaces have been defined and implemented for XML, CSV, HTML, DOC, DOCX, RTF, PDF and plain text data. Furthermore, the Text Corpus Format (TCF) is supported \cite{Heid10acorpus}. To process heterogeneous forms of text such as social media data (e.g. from Twitter or Facebook), administrative protocols, scientific publications or literary texts, a metadata management system is being introduced to define and manage corpus-specific metadata schemes. 
The LCM is designed to support multiple languages. We provide direct support for German and English text data. For other languages, a detailed documentation will be provided on how to customize configurations and resources (e.g. training models, stop word lists, lemma dictionaries) to add support for these languages.

\paragraph{Document Search:} With a full-text search and metadata fields integrated into the LCM, documents can be filtered by keywords and metadata facets. A query language enables the complex combination of multiple search criteria, such as AND/OR combinations of search terms, the exclusion of certain terms, or the search for terms which must occur within a certain minimum distance from each other. 
Search result sets can be stored as individual collections which can serve as a basis for a refined search or for any further text mining analysis. Results from such analysis can then be aggregated with respect to the collection level. Moreover, each document can be displayed in a full-text view to allow for a close reading and qualitative checks of text mining results.

\subsection{Analysis Features}

During the previous ePol project and from the experiences of third parties using the LCM prototype, functional requirements were identified that promise to significantly increase the added value of using the LCM for social science projects. These functions are integrated into the new implementation of the iLCM infrastructure in a modular manner. Modularity is ensured by standardized wrappers around existing text analysis functions, mostly provided as R libraries. By relying on a well-defined wrapper layer around those pre-existing functions, we allow for an easy integration of different analysis packages of the same kind (e.g. different approaches of key term extraction, topic models or text classification) into the analysis process management and visualization provided by the LCM. In the following, we introduce the main modules.

\paragraph{Linguistic (pre-)processing:}
For most text mining applications, text data must be transformed into a numeric representation. For instance, word counts per document over the entire vocabulary can be represented as a vector. All document vectors of a collection together form a document-term matrix (DTM) on which various text statistical evaluations can be processed. To generate such a DTM representation of a collection, we need several preprocessing capabilities. The LCM provides sentence segmentation, multi-word expression detection, POS-Tagging, Named Entity Tagging, Named Entity unification, lowercasing, stemming, lemmatization, stop word removal, ngram tokenization and data pruning which can be applied to the text data as needed. Additionally, the process chain for the linguistic preprocessing of documents is extended by syntactic parsing which is needed for an extended text analysis (e.g. subject-object relationships).  To fulfill those requirements we integrate the spaCy library. The open source Python library provides state of the art machine learning for natural language processing in multiple languages. 
Although spaCy is written in Python it can easily be integrated into the R environment with additional R packages \cite{Arnold.2017,Benoit.2018}.

\begin{figure}[t]
\begin{center}
\includegraphics[scale=0.125]{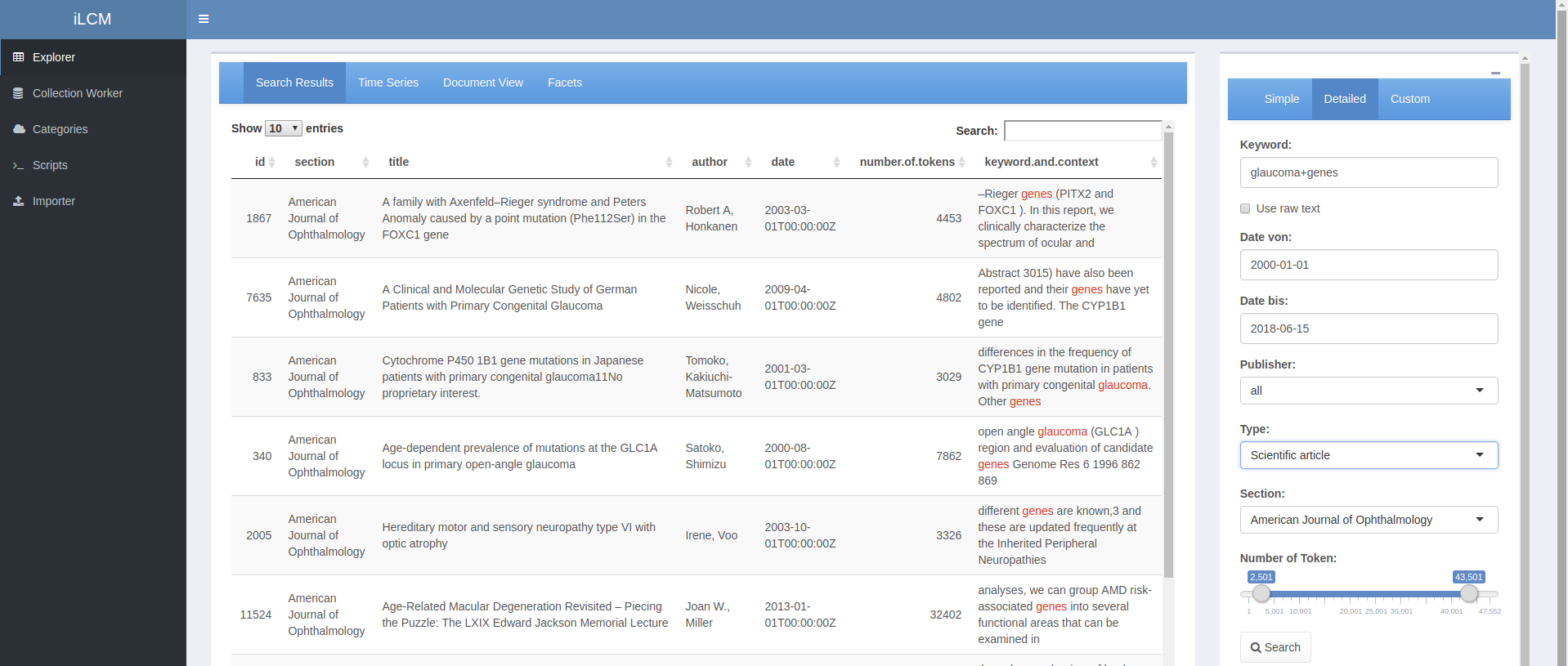} 
\caption{\textbf{LCM Document Search.} Screenshot of the document search interface.}
\label{iLCM-ds}
\end{center}
\end{figure}

\paragraph{Faceted Statistics:} Result sets of search queries can be filtered with respect to certain indexed metadata. With this, a user can quickly generate a graphical representation of the document metadata for a search query or create time series for the search results. The occurrence of proper names in query related texts can also be displayed in an aggregated form which quickly allows for an overview of entities that can be identified with respect to a search query.

\paragraph{Term Extraction/Keyword in Context:} Lists of meaningful terms can be calculated by the aggregation of their probabilities in different topics produced by a topic model. Those terms can be used to characterize the content of large collections or for further steps such as the creation of dictionaries or the identification of candidates for individual term analyses. Furthermore, terms can be displayed as "Keyword in Context" (KWIC).

\paragraph{Word Frequencies/Dictionary Analysis:} Similar to the measurement of document frequencies in relation to search results, the software provides possibilities for measuring term and document frequencies in document collections. According to the settings of a linguistic preprocessing chain, the articles in a collection are tokenized and the tokens are then counted, e.g. with reference to the publication dates of the articles. This allows the creation of time series for terms in absolute or normalized manner. Instead of single terms, also groups of terms representing semantically similar concepts, so-called dictionaries, can be used for frequency analysis.

\paragraph{Co-occurrence Analysis:} For the identification of semantic correlations, the statistical significance of word co-occurrences is decisive in addition to their frequency. We integrate several ways to compute the statistical significance of the co-occurrence of tokens within defined contextual units. Possible contextual units are sentences, paragraphs or documents. Furthermore, it is possible to extract co-occurrences from contextual windows (e.g. n-left and n-right) and to analyse the ordering of the co-occurring words in the contextual units. 

\begin{figure}[t]
\begin{center}
\includegraphics[scale=0.125]{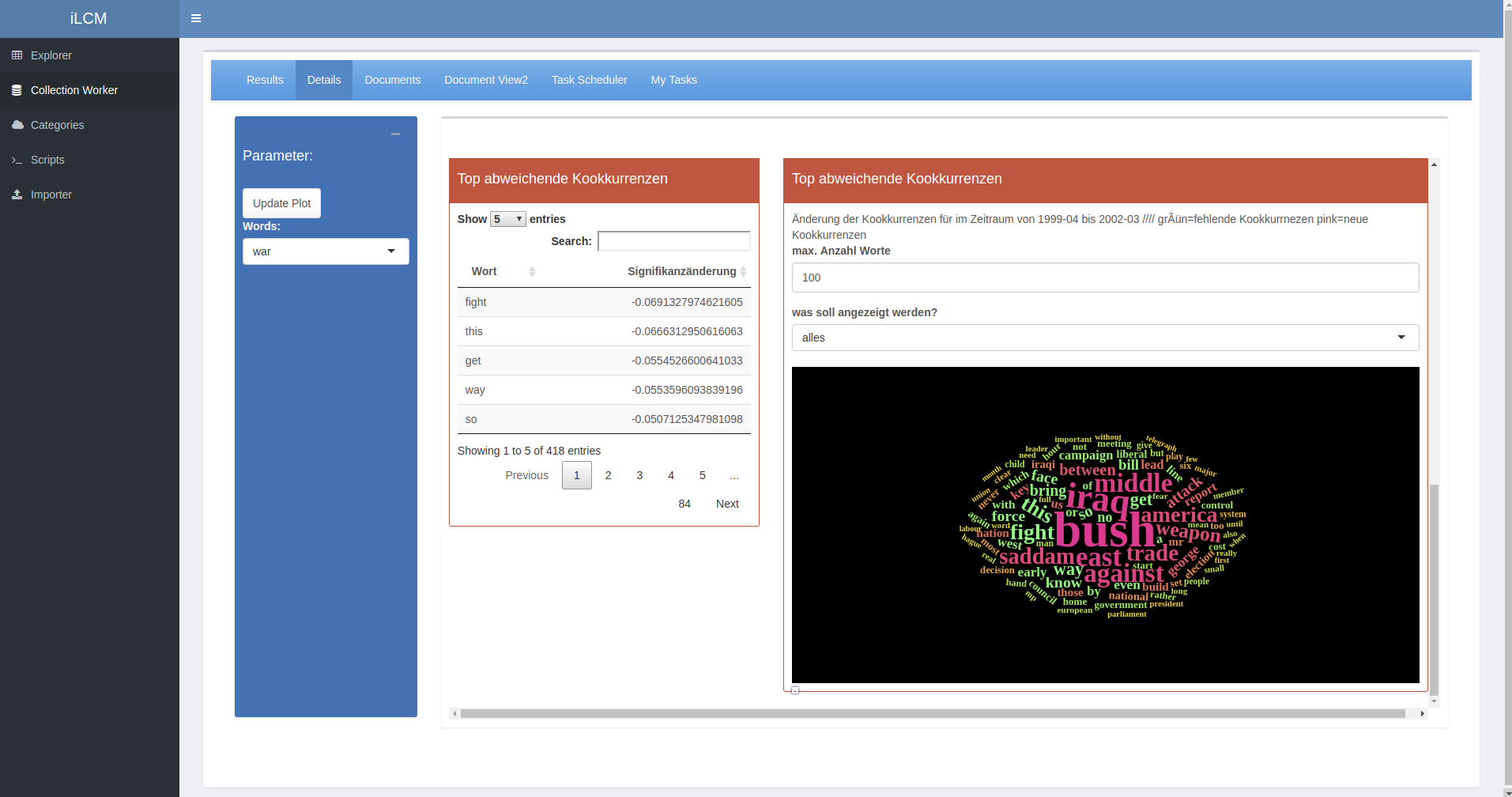} 
\caption{\textbf{LCM Context Volatility Results.} Screenshot of context volatility visualization.}
\label{iLCM-cv}
\end{center}
\end{figure}

\paragraph{Context Volatility:}
Using the LCM enables researchers to deal with diachronic text data. The amount of contextual change for certain words over a period of time can be analyzed using a measure which captures this property \cite{KNH17,Lietz.2014}. In the LCM several settings of measuring context change along with state of the art visualizations for the results are be implemented (cp. Fig.~\ref{iLCM-cv}).  

\begin{figure}[t]
\begin{center}
\includegraphics[scale=0.125]{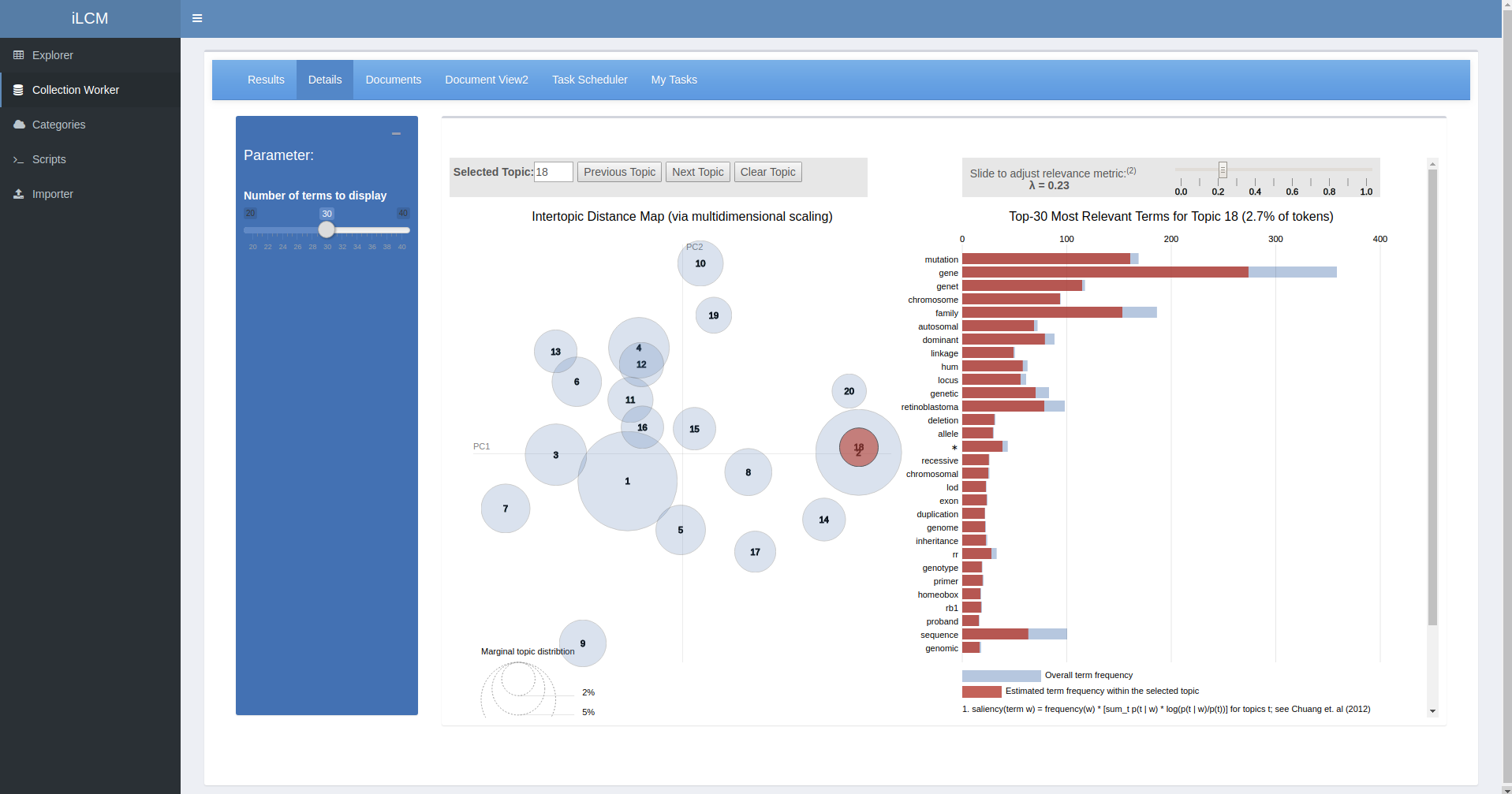} 
\caption{\textbf{LCM Topic Model Results.} Screenshot of a topic model visualization.}
\label{iLCM-tm}
\end{center}
\end{figure}

\paragraph{Topic Models:} Topic models are probabilistic models that attempt to infer semantic clusters (which can be interpreted as topics) in document collections. As a result, topic models provide probability distributions over the set of all words for each topic \cite{Blei2012}. Furthermore, they provide probability distributions over the set of topics for each document. A topic can be semantically interpreted using the $n$ most probable words it contains. For this purpose, the software provides sorted term lists for each calculated topic as well as an integration of the topic model result browser LDAvis \cite{Sievert.2014} (cp. Fig.~\ref{iLCM-tm}). We further add functionality to determine optimal hyper-parameters for the modeling process and to measure its reliability over repeated runs \cite{Maier.2018}. 

\paragraph{Manual Annotation:} Manual methods of qualitative data analysis usually work with categories. Those could be obtained inductively from the empirical material through reading and interpretation. More importantly, categories can be deductively guided and operationalized on the basis of existing theories (e.g. with the help of lists of terms, so-called dictionaries). This is supported by the definition of hierarchical category systems, encoding of entire documents or text segments with categories of the category system, measurement of intercoder matches and administration and export of the manually coded data. Due to the browser-based user interface, several annotators can simultaneously work on the same document set with a shared category system.

\paragraph{Supervised Classification:}
Categories for which a high intercoder reliability can be achieved, i.e. which are defined sufficiently unambiguously so that different encoders assign (mostly) the same categories for documents or passages of text, are also suitable for automatic classification procedures. Supervised methods for classification use given training examples to learn how to classify a data object into a defined class or category. Accordingly, the LCM offers the possibility of accepting text passages assigned to category X as positive training examples. Text passages in annotated documents that are not coded with X become negative training examples. A machine classifier learns textual features (especially word occurrences and their combinations) which suggest the existence of a certain class. This allows for an active learning scenario, where new text passages fitting to a certain category are identified within a collection automatically \cite{Wiedemann.2018}. In an iterated process, automatic suggestions can be corrected by a human annotator to improve the classification model.
In addition to text (passage) categorization, we further provide functionality for sequence classification to allow for complex information extraction.

\subsection{Analysis workflows} 

The individual procedures already allow for interesting analysis of large text collections. But, in a CSS scenario, most benefit is to be expected from combined applications of the analysis modules in complex workflows.
For instance, a topic model result is useful for interpretation in itself already. However, for content analysis, it is also interesting to use the topic model result as a filter for the original document collection. The result could be used to split a heterogeneous collection into thematically coherent sub-collections. These filtered sub-collections can be analyzed further as time series on the prominence of a theme. Also, further analysis procedures such as dictionary analysis could be applied to test hypotheses with respect to certain topics.
In another scenario, frequency observations over time can be used to identify periods that indicate a similar use of terms and points in time at which changes occur (e.g. peaks in time curves). These periods of time can be used to divide a collection into distinct sub-collections on which further comparative analyses such as co-occurrence analysis can be applied. 
In summary, the LCM allows for various combinations of the provided modules in creative ways which fit the research requirements of the end-users \cite{Lemke2015}. Information about the source collection, the selected analysis process and corresponding parameters for each analysis result are logged by the system in a way that allows for documentation of the conducted research.

\section{Open Research Computing}
\label{sec:orc}

The GUI-based framework of the LCM enables end-users to conduct standardized text mining workflows without any programming skills. However, individual and innovative research designs often demand more flexibility which is hard to achieve with generic pre-defined workflows accessed by point-and-click GUIs. Instead, such research designs can be supported well with script programming languages. The iLCM infrastructure integrates an Open Research Computing (ORC) environment to allow for the extension of LCM's generic analysis procedures by program scripts. In this way, the iLCM targets requirements of two user groups: beginners of text mining methods who benefit most from GUI-guided workflows as well as more advanced analysts who demand the flexibility of freely customizable scripts.

The ORC component is based on established open-source software such as Jupyter \cite{soton403913}, Docker and Kubernetes\footnote{\url{http://kubernetes.io}}, which are adapted to the requirements of the iLCM infrastructure. The development of these open-source frameworks is widely supported and hence provides the reliability of a well-maintained code base.

\subsection{Notebooks}
The ORC component extends the iLCM with an editor environment for program scripts which is operated via a web browser. The web editor enables the creation of scripts along with their documentation. Furthermore, scripts can be directly executed and the produced results can be visualized as output in the same document. 
The execution of the scripts themselves does not take place in the browser but on a server. This allows the processing of large data objects when server-side resources (memory, CPU) are available.

To fulfill important requirements for CSS, scripts can be extensively documented with markup code within the editor. Results from script executions such as numerical measures, tables or plots can be embedded in the same script document which allows tracing every single step of an analysis. Such actively executable documents, a combination of script code, its documentation, and corresponding results, are called ``notebooks''. For researchers this allows having the data, analysis scripts and their documentation available in one notebook for publication, sharing and re-use.

The ORC component extends the iLCM in two ways. First, the notebook environment is embedded in a virtual machine ensemble together with the LCM to closely integrate the GUI-paradigm for text analysis and the script paradigm to analyse structured data. Second, an ORC environment is publicly hosted by Gesis, the infrastructure partner of the iLCM project to allow for archiving, sharing and re-using of notebooks.

Various options are available to implement the notebook functionality, most notably R Notebooks and Jupyter Notebooks. For the iLCM, we opted for Jupyter Notebooks \cite{soton403913}. Jupyter Notebooks have a wide community support and enable the use of numerous scripting languages, including R, Python, and Julia.

\subsection{Notebook Gallery}
Gesis will host a public instance of the ORC environment. This ORC instance is then extended with a repository on which notebooks can be published by users. The repository not only secures the long-term availability of published notebooks, it also serves as a notebook gallery that users can browse and search via keywords and metadata. The notebooks can be shared, edited and executed in a SaaS manner. The execution of shared notebooks is enabled by the import of the notebooks into the ORC environment by a process referred to as ``cloning''. This supports and enhances the reproducibility of research in the field of CSS. 

The notebook gallery supports the research process in a number of ways. 
First of all, the gallery allows researchers to not only easily reproduce results of other researchers, but it also makes it easier to build on existing research through the cloning of notebooks. Furthermore, generic versions of different methods and analysis processes will be published as notebooks, enabling researchers to clone them and adapt them according to their own needs. Finally, the gallery makes it easy to discover learning materials published as notebooks, which supports researchers in learning new methods and analysis techniques.

We expect that the synergetic use of script-based analyses and generic text processing will support researchers in the development of analysis processes according to their project-specific requirements. We also expect a high impact of the architecture for teaching CSS methods.

\section{Integration}
\label{sec:integration}

The LCM is optimized for the generic processing of large amounts of text data. However, for experienced researchers, it is easy to identify needs for analyzes that go beyond the generic usage of text mining. The combination of standardized computer-linguistic evaluations in the LCM and customized evaluations in the ORC environment is an elegant solution. Therefore, the ORC environment is integrated into the iLCM architecture. Interfaces and functions will be implemented in the iLCM to easily exchange data objects between the LCM and the ORC environment (see Figure \ref{fig:architecture}). 
For instance, output from the LCM component for further processing in the ORC environment can be entire text corpora, previously selected text collections, linguistically preprocessed DTMs, topic model results or any other information extracted from texts by the standardized LCM procedures. 

In addition to further processing of LCM data objects, external data objects can be imported to combine complex analysis of both, structured and unstructured data.
The integration of external resources offers considerable added value. The European infrastructure CLARIN hosts particularly valuable text corpora \cite{hinrichs_krauwer_clarin_2014}. In addition, central search functionalities (Virtual Language Observatory, Federated Content Search) are available for the resources hosted in CLARIN, with which (sub-)corpora can be easily retrieved and processed in iLCM infrastructure. Interfaces defined by CLARIN for data import and export will be implemented in the iLCM. 
To facilitate the use of external resources with restricted access rights, iLCM will integrate a user rights management based on the Shibboleth\footnote{\url{http://shibboleth.net}} authentication and authorization. The rights management can be easily connected to identity providers such as managed by Gesis or by the CLARIN infrastructure to grant access for selected data objects and analysis to selected users only.

\subsection{Expected Impact}

We expect that the iLCM infrastructure will lead to a significant strengthening of the emerging field of CSS and the paradigm of reproducible research. The focus on reproducible research also goes hand in hand with the strengthening of the open access concept, since the publication of notebooks as descriptions of the research process is usually closely linked to the publication of the final manuscript of a research paper. The public availability of notebooks facilitates the review process and will eventually increase the chance to publish a final research manuscript as open access. 

Another advantage for the digital humanities and the social sciences is the opportunity for interdisciplinary exchange. So far, the different disciplines have used computer-supported text analysis relatively independently of each other. In sociology, political science, media and communication sciences, (socio-)linguistics or historical studies, similar methods of text analysis have been used for content analysis or discourse research. There are only few examples of computer-supported text analyses that go beyond the boundaries of those disciplines. Since all disciplines are currently facing the challenge of integrating complex algorithmic evaluations into their research processes, we expect high benefits from cross-disciplinary methodological debates. The iLCM environment could be a hub to facilitate such debates by enabling scholars from different disciplines to easily share their methodical procedures.

Last but not least, the iLCM infrastructure is also ideal for teaching. On the one hand, it facilitates the transfer of basic skills in the field of text mining through the easy access to analysis methods with the GUI of the LCM. On the other hand, the infrastructure supports the teaching of more complex and individualized approaches to text mining by the ORC environment. The concept of notebooks further provides an interesting new format for term papers in which the student's methodical proceeding is documented comprehensibly and in a reproducible manner.

\subsection{Exemplary Use-Cases}
\label{sec:usecases}

The capabilities of the integrated environment allow for flexible and extensive analysis applications. In order to illustrate the possibilities, we describe three potential use-cases.

\paragraph{Gender biases in Wikipedia:} Wikipedia contains numerous biographies of important personalities. An examination of the contents of these biographies can clarify whether there are significant differences in the description of the biographies of male and female persons in contemporary history. A corpus of several thousand selected Wikipedia biographies, along with the gender of the personalities, could be imported into the LCM. In the LCM, two sub-collections, female and male biographies, are created and keywords are extracted from both. The respective complementary corpus serves as a comparative corpus. Furthermore, link structures between personalities of different genders can be investigated.These link structures can then be exported to the ORC environment.There, custom scripts can be created to further analyze the link network with respect to the question of gender biases.

\paragraph{Topic-person networks:} 
Consider the following research question: Are people from politics and people from business more likely to be mentioned together in some economic policy contexts than in others? For answering this question, we analyze a collection of newspaper articles on economic policy using the iLCM infrastructure. First, the LCM component can be used for a named entity extraction to generate a co-occurrence network of politicians and business leaders. For the assignment of people to the category economics or politics, a link to a corresponding resource database such as DBPedia\footnote{\url{http://dbpedia.org}} can be used. Furthermore, the LCM component can be used to compute a topic model on the text of the news articles. The results of both computations can then be imported into the ORC environment for a more fine-grained analysis of the people network with respect to the topical contexts they co-occur in.

\paragraph{CSS teaching lab:} 
Students of CSS can be taught to use state-of-the-art text mining with the help of the iLCM infrastructure. The GUI-based LCM component provides an easy and comprehensible access to basic analysis methods on large text collections such as full corpora of newspapers. In several in-class exercises, students are introduced to single methods such as key term extraction or topic modeling. With the help of the retrieval function, they are requested to select a suitable subcorpus from a full corpus of newspaper articles (e.g. NYTimes corpus) with respect to a given research question. Then, they apply the specific analysis method, try out different process parameters and in the end discuss the results in class. Since results may vary due to different initial collection retrieval processes or parameter settings, students will learn about the effects of each of the process steps by comparing their results. At the end of each term, students are requested to hand in a notebook created in the ORC component with an own study on the corpus used during the in-class exercises.

\section{Conclusion}
\label{sec:conclusion}

In this paper, we presented the iLCM, a SaaS infrastructure comprising a component for the analysis of large unstructured text collections (Leipzig Corpus Miner, LCM) and a component for the (statistical) analysis of structured data (Open Research Computing, ORC). The combination of both allows for sophisticated computer-assisted text analysis in the context of computational social science and digital humanities research. It further fulfills requirements from the targeted user audience regarding different levels of previous knowledge about computational methods and reproducible research. This is achieved by providing a GUI-based approach to text analysis (LCM) and an opportunity to code own analysis scripts (ORC). The decentralized iLCM architecture is accompanied by a central repository of so-called notebooks, analysis scripts along with data, processed results and a process documentation to allow for archiving, sharing and re-use of text analysis workflows.
By this, the iLCM enables social scientists to work with large amounts of data in order to produce pioneering results. 

Requirements for the infrastructure were derived from experiences in a previous project in which a prototype of the LCM was implemented and used for several social science studies. For the current project, we plan to conduct additional user studies to learn more about the requirements of the targeted disciplines, to improve the usability of the GUI and to extend the text analysis capabilities. At the time of publication of this paper, the work on the iLCM is still in progress. Hence, the described environment is not feature complete yet. Nonetheless, an early version is already available on Github and will be developed further in our public repository.

Our future work concentrates on the completion of the environment, sufficient documentation, open-source releases of the developed components and user testing in order to create a sustainable and convenient research environment for the social sciences and the humanities.

\section{Acknowledgements}

The project iLCM is funded by the German Research Foundation (DFG).  FKZ/project number: 324867496. 

\section{Bibliographical References}
\label{main:ref}
\bibliographystyle{lrec}
\bibliography{references}

\end{document}